\documentclass[11pt]{article}
\usepackage{color}
\usepackage{amssymb,latexsym,bm, amsmath,amsthm, amsfonts,multirow,graphicx,mathrsfs}
\usepackage{eufrak,booktabs}
\usepackage{epsfig}
\usepackage{enumitem}
\graphicspath{{figures/}}

\newcommand{\argmax}{\mathop{\rm argmax}}

\def\Deltabf{{\boldsymbol      \Delta}}

\def\Ab{{\boldsymbol A}}
\def\Bb{{\boldsymbol B}}

\def\Hb{{\boldsymbol H}}
\def\Ib{{\boldsymbol I}}
\def\Jb{{\boldsymbol J}}
\def\Kb{{\boldsymbol K}}

\def\Mb{{\boldsymbol M}}

\def\Pb{{\boldsymbol P}}

\def\Ub{{\boldsymbol U}}
\def\Vb{{\boldsymbol V}}
\def\Wb{{\boldsymbol W}}
\def\Xb{{\boldsymbol X}}
\def\Yb{{\boldsymbol Y}}

\def\etab{{\boldsymbol        \eta}}

\def\Lambdab{{\boldsymbol     \Lambda}}

\def\Sigmab{{\boldsymbol      \Sigma}}

\def\vec{{\rm vec}}

\def\vecp{{\rm vecp}}
\def\1{{\bm 1}}
\def\0{{\bm 0}}
\def\th{{\rm th}}

\def\Xb{{\bm X}}\def\Wb{{\bm W}}
\def\rmSC{{\rm Screen-and-Clean~}}
\def\rmESC{{\rm Extended Screen-and-Clean~}}

\def\CI{{\cal I}}

\newcommand{\CS}{\mathcal{S}}

\newcommand{\E}{\mathcal{E}}

\newtheorem{thm}{Theorem}

\newtheorem{rmk}[thm]{Remark}

\begin{document}

\title{Detection of Gene-Gene Interactions by Multistage\\
Sparse and Low-Rank Regression}
\date{\empty}

\author{Hung~Hung$^a$, Yu-Tin Lin$^b$, Pengwen~Chen$^c$, Chen-Chien Wang$^d$\\ Su-Yun~Huang$^b$, and Jung-Ying Tzeng$^e$\footnote{To whom
correspondence should be addressed. E-mail: \textit{jytzeng@ncsu.edu}}\\[2ex]
$^a$Institute of Epidemiology \& Preventive Medicine\\[-1ex] National Taiwan University\\[1ex]
$^b$Institute of Statistical Science, Academia Sinica\\[1ex]
$^c$Department of Applied Mathematics, National Chung
Hsing University\\[1ex]
$^d$Department of Computer Science, New York University\\[1ex]
$^e$Department of Statistics and Bioinformatics Research Center\\[-1ex]
 North Carolina State University}

\maketitle

\begin{abstract}

A daunting challenge faced by modern biological sciences is finding
an efficient and computationally feasible approach to deal with the
curse of high dimensionality. The problem becomes even more severe
when the research focus is on interactions. To improve the
performance, we propose a low-rank interaction model, where the
interaction effects are modeled using a low-rank matrix. With
parsimonious parameterization of interactions, the proposed model
increases the stability and efficiency of statistical analysis.
Built upon the low-rank model, we further propose an Extended
Screen-and-Clean approach, based on the Screen and Clean (SC) method
(Wasserman and Roeder, 2009; Wu \textit{et al}., 2010), to detect
gene-gene interactions. In particular, the screening stage utilizes
a combination of a low-rank structure and a sparsity constraint in
order to achieve higher power and higher selection-consistency
probability. We demonstrate the effectiveness of the method using
simulations and apply the proposed procedure on the warfarin dosage
study. The data analysis identified main and interaction effects
that would have been neglected using conventional methods.\\

%Individual hypothesis tests are carried out on each interaction
%effect selected in the screening stage for a final model.

%{\color{red} We have implemented a xxx algorithm for the proposed
%procedure and derived related asymptotic distribution}.

\end{abstract}

\section{Introduction}\label{sec.1}

Modern biological researches deal with high-throughput data and encounter the curse of
high-dimensionality. The problem is further exacerbated when the question of interest focuses on gene-gene interactions (G$\times$G).
Due to the extremely high-dimensionality for modeling G$\times$G, many G$\times$G
methods are multi-staged in nature that rely on a screening step to reduce the
number of loci (Cordell 2009; Wu et al. 2010). Joint screening based on the multi-locus
model with all main effect and interactions terms is preferred over marginal screening
based on single-locus tests --- it improves the ability to identify loci that interact
with each other but exhibit little marginal effect (Wan et al. 2010) and
improves the overall screening performance by reducing the unexplained variance
in the model (Wu et al. 2010). However, joint screening imposes statistical and
computational challenges due to the ultra-large number of variables. To tackle
this problem, one promising method that has good results is the Screen
and Clean (SC) procedure (Wasserman and Roeder, 2009; Wu et al. 2010). The SC
procedure first uses Lasso to pre-screen candidate loci where only main effects
are considered. Next, the expanded covariates are constructed to include the
selected loci and their corresponding pairwise interactions, and
another Lasso is applied to identity important terms. Finally, in the cleaning stage with an
independent data set, the effects of the selected terms are estimated
by least squares estimate (LSE) method, and those terms that
pass $t$-test cleaning are identified to form the final model.

A crucial component of the SC procedure is the Lasso step in the screening process for interactions. Let $Y$ be the
response of interest and $G=(g_1,\cdots,g_p)^T$ be the genotypes at
the $p$ loci. A typical model, which is also the model considered in SC, for G$\times$G detection is
\begin{eqnarray}
E(Y|G)= \gamma +\sum_{j=1}^p\xi_j\cdot g_j+  \sum_{j<
k}\eta_{jk}\cdot (g_j\,g_k),
\label{model_full0}
\end{eqnarray}
where $\xi_j$ is the main effect of the $j^\th$ loci, and
$\eta_{jk}$, $j< k$, is the G$\times$G corresponding to the $j^\th$
and $k^\th$ loci.
The Lasso step of SC then fits model~(\ref{model_full0}) to reduce the model size from
\begin{eqnarray}
m_p=1+p+{p\choose2}\label{mp}
\end{eqnarray}
to a number relatively smaller than sample size, $n$, based on which the validity of
the subsequent LSE cleaning can be guaranteed.
%%
%% If the screening is
%% restrictive, effective covariates cannot be identified, and the
%% procedure will fail to be selection consistent. On the other hand,
%% if too many covariates survive at this stage, insignificant
%% covariates will be included after the screening stage, and the
%% $t$-tests in the cleaning stage will have low powers.
%%
The performance of Lasso is known to depend on the involved number
of parameters $m_p$ and the available sample size $n$.  Although
Lasso has been verified to perform well for large $m_p$, caution
should be used when $m_p$ is ultra-large such as in the order of
$\exp\{O(n^\delta)\}$ for some $\delta>0$ (Fan and Lv, 2008). In
addition, the $m_p$ encountered in modern biomedical study is
usually greatly larger than $n$ even for a moderate size of $p$. In
this situation, statistical inferences can become unstable and
inefficient, which would impact the screening performance and
consequently affect the selection-consistency of the SC procedure or
reduce the power in the $t$-tests cleaning.

To improve the exhaustive screening involving all main and interaction terms, we consider a
reduced model by utilizing the matrix nature of interaction terms.
Observing model~(\ref{model_full0}) that $(g_j\,g_k)$ is the $(j,k)^\th$ element of the
symmetric matrix $\Jb=GG^T$, it is natural to treat $\eta_{jk}$ as
the $(j,k)^\th$ entry of the symmetric matrix $\etab$, which leads
to an equivalent expression of model~(\ref{model_full0}) as
\begin{eqnarray}
E(Y|G) =\gamma+\xi^T G +\vecp(\etab)^T \vecp(\Jb),\label{model_full}
\end{eqnarray}
where $\xi=(\xi_1,\ldots,\xi_p)^T$ and $\vecp(\cdot)$ denotes the
operator that stacks the lower half (excluding diagonals) of a
symmetric matrix columnwisely into a long vector. With the model
expression~(\ref{model_full}),
%% the interaction parameter of interest is the symmetric matrix $\etab$, and
we can utilize the structure of the symmetric matrix $\etab$ to improve the inference
procedure. Specifically, we posit the condition for the
interaction parameters
\begin{equation}\label{sparse_LR}
\etab: \mbox{being sparse and low-rank}.
\end{equation}
Condition~(\ref{sparse_LR}) is typically satisfied in modern
biomedical research. First, in a G$\times$G scan, it is reasonable
to assume most elements of $\etab$ are zeros because only a small
portion of the terms are related to the response $Y$. {This sparsity
assumption is also the underlying rationale for applying Lasso for
}variable selection in conventional approaches (e.g., Wu's SC
procedure). Second, if the elements of $\etab$ are sparse, the matrix $\etab$ is also likely
to be low-rank. Displayed below is an example
of $\etab$ with $p=10$ that contains three pairs of non-zero
interactions, and hence has rank 3 only:
\begin{equation}\label{eta_structure}
\etab=\left[\begin{array}{cc}\begin{array}{ccc}
0& \bigstar &\spadesuit\\
 \bigstar&0&\blacklozenge\\
\spadesuit&\blacklozenge&0\\
\end{array}&\0_{3\times 7} \\ \0_{7\times 3}&\0_{7\times7}\end{array}\right].
\end{equation}

One key characteristic in our proposed method is the consideration
of the sparse and low-rank condition~(\ref{sparse_LR}), which allows
us to express $\etab$ with much fewer parameters. In contrast,
Lasso does not utilize the matrix structure but only assumes the
sparsity of $\etab$ and, hence, still involves ${p\choose2}$
parameters in $\etab$. From a statistical viewpoint, parsimonious
parameterizations can improve the efficiency of model inferences.
Our aims of this work are thus twofold. First, {using
model~(\ref{model_full}) and  condition (\ref{sparse_LR}),
we propose an efficient screening procedure referred}
to as the sparse and low-rank screening (SLR-screening). Second,
{we demonstrate how the SLR-screening can be incorporated into existing multi-stage GxG methods to enhance the
power and selection-consistency. Based on the promise of the SC procedure, we illustrate the concept by proposing the \rmESC (ESC) procedure,
which replaces the Lasso screening with SLR-screening in the standard SC procedure.}

Some notation is defined here for reference. Let
$\{(Y_i,G_i)\}_{i=1}^n$ be random copies of $(Y,G)$, and let
$\Jb_i=G_iG_i^T$. Let $\Yb=(Y_1,\cdots,Y_n)^T$ be an $n$-vector of
observed responses, and let $\Xb=[X_1,\cdots,X_n]^T$ be the design
matrix with $X_i= [1, G_i^T, \vecp(\Jb_i)^T ]^T$. For any square
matrix $\Mb$, $\Mb^-$ is its Moore-Penrose generalized inverse.
$\vec(\cdot)$ is the operator that stacks a matrix columnwisely into a long vector.
$\Kb_{p,k}$ is the commutation matrix such that $\Kb_{p,k}\,
\vec(\Mb) = \vec(\Mb^T)$ for any $p\times k$ matrix $\Mb$ (Henderson
and Searle, 1979; Magnus and Neudecker, 1979). $\Pb$ is the matrix
satisfying $\Pb\vec(\Mb) = \vecp(\Mb)$ for any $p\times p$ symmetric
matrix $\Mb$. $\Pb$ can be chosen such that $\Pb\Kb_{p,p}=\Pb$. For a
vector, $\|\cdot\|$ is its Euclidean norm (2-norm), and
$\|\cdot\|_1$ is its 1-norm. For a set, $|\cdot|$ denotes its
cardinality.

\section{Inference Procedure for Low-Rank Model}\label{sec.low.rank}

\subsection{Model specification and estimation}\label{low.rank.fit}

To incorporate the low-rank property (\ref{sparse_LR})
into model building, for a pre-specified positive integer $r\le p$,
we consider the {following} rank-$r$ model
\begin{equation}
E(Y|G) = \gamma + \xi^T G +\vecp(\etab)^T \vecp(\Jb),\quad{\rm
rank}(\etab)\le r.\label{model}
\end{equation}
Although the above low-rank model expression is straightforward, it
is not convenient for numerical implementation. In view of this
point, we adopt an equivalent parameterization $\etab(\phi)$ for
$\etab$ that directly satisfies the constraint rank$(\etab)\le r$.
Consider the case with the minimum rank $r=1$ (the rank-1 model), we
use the parameterization
\begin{equation}
\etab(\phi)=u\alpha\alpha^T,\quad\phi=(\alpha^T,u)^T,\quad\alpha\in
\mathbb{R}^p, \quad u\in \mathbb{R}. \label{phi_rank1}
\end{equation}
For the case of higher rank, we consider the parameterization
\begin{eqnarray}
\etab(\phi) = \Ab\Bb^T + \Bb\Ab^T,\quad \phi=\vec(\Ab,\Bb)^T,\quad
\Ab,\Bb\in \mathbb{R}^{p\times k},  \label{phi_AB}
\end{eqnarray}
which gives $r=2k$ (the rank-$2k$ model), since the maximum rank
attainable by $\etab(\phi)$ in (\ref{phi_AB}) is $2k$.
Note that in either cases of (\ref{phi_rank1}) or (\ref{phi_AB}),
the number of parameters required for interactions $\etab(\phi)$ can be largely
smaller than $p\choose2$. See Remark~\ref{rmk.over} for more
explications. Thus, when model~(\ref{model}) is true, standard MLE arguments
show that statistical inference based on model~(\ref{model})
must be the most efficient. Even if model~(\ref{model}) is incorrectly
specified, when the sample size is small, we are still in favor of the low-rank model.
In this situation, model~(\ref{model}) provides a good ``working''
model. It compromises between the model
approximation bias and the efficiency of parameters estimation. With limited sample size,
instead of unstably estimating the full model, it is preferable
to more efficiently estimate the approximated low-rank model.
As will be shown later, a low-rank approximation of $\etab$ with
parsimonious parameterization suffices to more efficiently screen out relevant interactions.

%One advantage of \eqref{phi_rank1} is the parsimony of parameters
%used. It can be seen that we only need $(p+1)$ parameters
%$\phi=(\alpha^T,u)^T$ for modeling interactions, while it is
%$p(p-1)/2$ for full model~(\ref{model_full}).

%Let $q_r$ be the number of parameters required to specify
%model~(\ref{model}). In this article we will assume $q_r\ll n\ll
%m_p$ when fitting a rank-$r$ model. That is, we are able to fit
%model~(\ref{model}) efficiently, although this is not the case for
%the full model~(\ref{model_full}).

Let the parameters of interest in the rank-$r$ model~(\ref{model}) be
\begin{equation}\label{b_theta}
\beta(\theta)=\left[\gamma,\xi^T,\vecp\{\etab(\phi)\}^T\right]^T\quad{\rm
with}\quad\theta=\left(\gamma,\xi^T,\phi^T\right)^T,
\end{equation}
which consist of intercept, main effects, and interactions. Under
model~(\ref{model}) and assuming i.i.d. errors from a normal
distribution $N(0,\sigma^2)$, the log-likelihood function (apart
from constant term) is derived to be
\begin{eqnarray}
\ell(\theta)=-\frac{1}{2}\sum_{i=1}^n\left\{Y_i-\gamma -\xi^T G_i -
\vecp\{\etab(\phi)\}^T\vecp(\Jb_i)\right\}^2=-\frac{1}{2}\|Y-\Xb
\beta(\theta)\|^2.
\end{eqnarray}
To further stabilize the maximum likelihood estimation {MLE}, a
common approach is to append a penalty on $\theta$ to the
log-likelihood function. We then propose to estimate $\theta$
through maximizing the penalized log-likelihood function
\begin{align}
\ell_{\lambda_\ell}(\theta) = \ell(\theta)- \frac{\lambda_{\ell}}2\,
\| \theta \|^2,\label{criterion}
\end{align}
where $\lambda_{\ell}$ is the penalty (the subscript $\ell$ is for
low-rank). Denote the penalized MLE as
\begin{eqnarray}
\widehat\theta_{\lambda_{\ell}}=
\left(\widehat\gamma_{\lambda_{\ell}},\, \widehat\xi_{\lambda_{\ell}},\,
\widehat\phi_{\lambda_\ell}^T\right)^T
=\argmax_{\theta}\,\ell_{\lambda_\ell}(\theta).\label{mle.theta}
\end{eqnarray}
The parameters of interest $\beta(\theta)$ are then estimated by
\begin{eqnarray}
\widehat\beta_{\lambda_\ell}=\beta(\widehat\theta_{\lambda_{\ell}}),\label{est.beta}
\end{eqnarray}
on which subsequent analysis for main and G$\times$G effects can be based.
In practical implementation, we use $K$-fold cross-validation
($K=10$ in this work) to select $\lambda_\ell$.

%To pick out significant main effects and interactions, we will do a
%series of entry-wise $t$-tests to construct $\CI_1$, the index set
%for  the selected model in the first stage. Detailed testing
%procedure and the asymptotic distribution of
%$\widehat\beta_{\lambda_\ell}$ necessary to carry out the hypothesis
%tests are discussed in Section~\ref{low.rank.asymptotics}.

\begin{rmk}\label{rmk.over}
We only need $pr-r^2/2+r/2$ parameters to specify a $p\times p$
rank-$r$ symmetric matrix, and the number of parameters required for
model~(\ref{model}) is
\begin{eqnarray}
d_r=1+p+(pr-r^2/2+r/2).\label{dr}
\end{eqnarray}
However, adding constraints makes no difference to our inference
procedures, but only increases the difficulty in computation. For
convenience, we keep this simple usage of $\phi$ without imposing
any identifiability constraint.
\end{rmk}

\subsection{Implementation algorithm}

\subsubsection{The case of rank-1 model}

For the rank-1 model $\etab(\phi)=u\alpha\alpha^T$, it suffices to
maximize (\ref{criterion}) using Newton method under both
$u=+1$ and $u=-1$. The one from $u=\pm1$ with the larger value of
penalized log-likelihood will be used as the estimate of $\theta$.
For any fixed $u$, maximizing (\ref{criterion}) is
equivalent to the minimization problem:
\begin{align}
\min_{\theta_u}\, \frac{1}{2} \left\|Y - \Xb_u
\beta_u(\theta_u)\right\|^2 + \frac{\lambda_{\ell}}{2} \big\|
\theta_u\big\|^2,\label{criterion.rank1}
\end{align}
where $\Xb_u = [ X_{u1}, \cdots,  X_{un}]^T$ with $X_{ui} = [1,
G_i^T, u\cdot\vecp(\Jb_i)^T]^T$ is the design matrix, and $\beta_u(\theta_u)=\{\gamma , \xi^T,\vecp(\alpha\alpha^T)^T\}^T$
with $\theta_u = (\gamma , \xi^T,
\alpha^T)^T$. Define
\begin{eqnarray*}
\Wb_u(\theta_u)=\Xb_u\,\frac{\partial
\beta_u(\theta_u)}{\partial\theta_u}\quad{\rm with}\quad
   \frac{\partial \beta_u(\theta_u)}{\partial\theta_u}=
   \left[\begin{array}{cc}
     \Ib_{p+1}&\0\\
     \0& 2\Pb(\alpha\otimes\Ib_p).
   \end{array}\right].
\end{eqnarray*}
The gradient and Hessian matrix (ignoring the zero expectation term)
of (\ref{criterion.rank1}) are
\begin{eqnarray*}
  g_u(\theta_u)&=&-\{\Wb_u(\theta_u)\}^T\{Y-\Xb_u \beta_u(\theta_u)\}+\lambda_{\ell}\theta_u,\\
  \Hb_u(\theta_u)&=&\{\Wb_u(\theta_u)\}^T\{\Wb_u(\theta_u)\}+\lambda_{\ell} \Ib_{2p+1}.
\end{eqnarray*}
Then, given an initial $\theta_u^{(0)}$, the minimizer
$\widehat\theta_u$ of (\ref{criterion.rank1}) can be obtained
through the iteration
\begin{eqnarray}
\theta_u^{(t+1)}=\theta_u^{(t)}-\left\{\Hb_u(\theta_u^{(t)})
\right\}^{-1}g_u(\theta_u^{(t)}),~t=0,1,2,\ldots,\label{newton}
\end{eqnarray}
until convergence, and output $\widehat\theta_u=\theta_u^{(t+1)}$.
Let $u^*$ correspond to the optimal $u$ from $u=\pm1$. The final
estimate is defined to be
$\widehat\theta_{\lambda_\ell}=(\widehat\theta_{u^*}^T,u^*)^T$.

\subsubsection{The case of rank-$\bm{2k}$ model}

When $\etab(\phi)=\Ab\Bb^T+\Bb\Ab^T$, we use the alternating least
squares (ALS) method to maximize (\ref{criterion}). By fixing $\Ab$,
the problem of solving $\Bb$ becomes a standard penalized least squares problem.
This can be seen from
\begin{align*}
  \vecp(\Ab\Bb^T + \Bb\Ab^T)
  %&= \Pb\vec(\Ab\Bb^T + \Bb\Ab^T) = \Pb\left\{ \vec(\Ab\Bb^T) + \Kb_{pp} \vec(\Ab\Bb^T) \right\} \\
  &= 2\Pb \vec(\Ab\Bb^T) = 2\Pb(\Bb \otimes \Ib_p) \vec(\Ab),
\end{align*}
where the second equality holds by $\Pb\Kb_{p,p}=\Pb$. Hence, maximizing (\ref{criterion}) with fixed $\Bb$
is equivalent to the minimization problem:
\begin{align}
\min_{ \theta_\Bb} \frac{1}{2} \left\|\Yb - \Xb_\Bb
\theta_\Bb\right\|^2 + \frac{\lambda_{\ell}}{2} \big\|
\theta_\Bb\big\|^2,\label{max.B}
\end{align}
where $\Xb_\Bb = [ X_{\Bb 1}, \cdots, X_{\Bb n}]^T$ with $X_{\Bb i}
= [1, G_i^T, 2\vecp(\Jb_i)^T \Pb(\Bb \otimes \Ib_p)]^T$ being the
design matrix when $\Bb$ is fixed, and $\theta_\Bb =\left[\gamma ,
\xi^T, \vec(\Ab)^T\right]^T$. It can be seen that (\ref{max.B}) is
the penalized least squares problem with data design matrix
$\Xb_\Bb$ and parameters $\theta_\Bb$, which is solved by
\begin{equation}
  \widehat{\theta}_\Bb = \left( \Xb^T_\Bb  \Xb_\Bb + \lambda_{\ell} \Ib_{1+p+pk}\right)^{-1}
   \Xb_\Bb^T \Yb. \label{fixB}
\end{equation}
Similarly, the maximization problem with fixed $\Ab$ is equivalent
to the minimization problem
\begin{align*}
\min_{\theta_\Ab}\, \frac{1}{2} \left\|\Yb - { \Xb}_\Ab
\theta_\Ab\right\|^2 + \frac{\lambda_{\ell}}{2} \big\| \theta_\Ab
\big\|^2,
\end{align*}
where $\Xb_A = [ X_{1\Ab}, \cdots,  X_{n\Ab}]^T$ with $X_{i\Ab} =
[1, G_i^T, 2\vecp(\Jb_i)^T \Pb(\Ab \otimes \Ib_p)]^T$ being the
design matrix when $\Ab$ is fixed, and $\theta_\Ab =\left[\gamma ,
\xi^T, \vec(\Bb)^T\right]^T$. Thus, when $\Ab$ is fixed,
$\theta_\Ab$ is solved by
\begin{equation}
  \widehat{\theta}_\Ab = \left( \Xb^T_\Ab  \Xb_\Ab + \lambda_{\ell} \Ib_{1+p+pk}\right)^{-1}
   \Xb_\Ab^T \Yb. \label{fixA}
\end{equation}
The ALS algorithm then iteratively and alternatively changes the roles of $\Ab$ and $\Bb$
until convergence. Detailed algorithm is summarized below.\\

\hrule

~\

\noindent \textbf{Alternating Least Squares (ALS) Algorithm:}
\begin{enumerate}
\item
Set initial $\Bb^{(0)}$. For $t=0,1,2,\ldots,$
\begin{enumerate}
\item[(1)]
Fix $\Bb=\Bb^{(t)}$, obtain
$\widehat{\theta}_{\Bb^{(t)}}=\{\gamma^{(t)},\xi^{(t)},\vec(\Ab^{(t+1)})^T\}^T$
from (\ref{fixB}).
\item[(2)]
Fix $\Ab=\Ab^{(t+1)}$, obtain
$\widehat{\theta}_{\Ab^{(t+1)}}=\{\gamma^{(t+1)},\xi^{(t+1)},\vec(\Bb^{(t+1)})^T\}^T$
from (\ref{fixA}).
\end{enumerate}

\item Repeat Step-1 until convergence.
Output $(\gamma^{(t+1)},\xi^{(t+1)},\Ab^{(t+1)},\Bb^{(t+1)})$ to
form $\widehat\theta_{\lambda_\ell}$.
\end{enumerate}
\hrule \vspace{0.7cm}

\noindent
Note that the objective
function value increases in each iteration of the ALS algorithm. In addition, the
penalized log-likelihood function is bounded above by zero, which
ensures that the ALS algorithm converges to a stationary
point. We found in our numerical studies that a random initial
$\Bb^{(0)}$ will converge quickly and produce a good solution.

%When fixing $\Ab$ (or $\Bb$), maximizing (\ref{criterion}) becomes a least-squares problem for
%solving $\gamma,\xi$ and $\Bb$ (or $\Ab$), and

\subsection{Asymptotic properties}\label{low.rank.asymptotics}

This subsection devotes to derive the asymptotic distribution of $\widehat\beta_{\lambda_\ell}$
defined in \eqref{est.beta}, which is the core to
propose our SLR-screening in the next section. Assume that the parameter space
$\Theta$ of $\theta$ is bounded, open and connected, and define
$\Xi=\beta(\Theta)$ be the induced parameter space. Let
$\beta_0=\{\gamma_0,\xi_0^T,\vecp(\etab_0)^T\}^T$ be the true
parameter value of the low-rank model~(\ref{model}) and define
\begin{eqnarray}
\Deltabf(\theta)=\frac{\partial}{\partial\theta}\beta(\theta).
\label{delta}
\end{eqnarray}
We need the following regularity conditions for deriving asymptotic
properties.
\begin{enumerate}[label=(C\arabic{*}),ref=C\arabic{*}]
\item \label{C1} Assume $\beta_0=\beta(\theta_0)$ for some $\theta_0\in\Theta$.
\item \label{C2}
Assume that $\beta(\theta)$ is locally regular at $\theta_0$ in the
sense that $\Deltabf(\theta)$ has the same rank as
$\Deltabf(\theta_0)$ for all $\theta$ in a neighborhood
of $\theta_0$. Further assume that there exists neighborhoods ${\cal
U}$ and ${\cal V}$ of $\theta_0$ and $\beta_0$ such that
$\Xi\cap{\cal V}=\beta({\cal U})$.
\item \label{C3}
Let $\Vb_n=\frac{1}{n}\Xb^T\Xb$. Assume that $\Vb_n
\stackrel{p}{\to} \Vb_0$ and that $\Vb_0$ is strictly positive
definite.
\end{enumerate}
The main result is summarized in the following theorem.

\begin{thm}\label{thm.asymp_dist}
Assume model~\eqref{model} and conditions \eqref{C1}-\eqref{C3}.
Assume also
%$\Vb_n=(\Xb^T\Xb)/n \stackrel{p}{\to} \Vb_0>0$ and
$\lambda_{\ell}=o(\sqrt{n})$. Then, as $n\to\infty$, we have
\begin{eqnarray}\label{asymp_dist}
\sqrt{n}(\widehat\beta_{\lambda_\ell}-\beta_0)\stackrel{d}{\rightarrow}
N(0,\Sigmab_0),
\end{eqnarray}
where $\Sigmab_0=\sigma^2\Deltabf_0(\Deltabf_0^T
\Vb_0\Deltabf_0)^{-}\Deltabf_0^T$ with
$\Deltabf_0=\Deltabf(\theta_0)$.
\end{thm}

%Indeed, it is the parameter $\beta(\theta)$ which is our
%major interest, and the parameter $\theta$ is only used as a means
%of low-rank parametrization.

To estimate the asymptotic covariance
$\Sigmab_0$, we need to estimate $(\sigma^2,\Deltabf_0)$. The error
variance $\sigma^2$ can be naturally estimated by
\begin{eqnarray}
\widehat\sigma^2=\frac{\|\Yb-\Xb\widehat\beta_{\lambda_\ell}\|^2}{n-d_r},\label{sigma}
\end{eqnarray}
where $d_r$ is defined in (\ref{dr}). We propose to estimate
$\Deltabf_0$ by
$\widehat\Deltabf_0=\Deltabf(\widehat\theta_{\lambda_\ell})$.
Finally, the asymptotic covariance matrix in
Theorem~\ref{thm.asymp_dist} is estimated by
\begin{eqnarray}
\widehat\Sigmab_0=\widehat\sigma^2\widehat\Deltabf_0\left\{\Ub\left(\Lambdab+\frac{\lambda_\ell}{n}
\Ib_{d_r}\right)\Ub^T \right\}^-\widehat\Deltabf_0^T,\label{est.cov}
\end{eqnarray}
where $\Ub\Lambdab\Ub^T$ is the singular value decomposition of
$\widehat\Deltabf_0^T \Vb_n\widehat\Deltabf_0$, $\Lambdab\in
\mathbb{R}^{d_r\times d_r}$ is the diagonal matrix consisting of
$d_r$ nonzero singular values with the corresponding singular
vectors in $\Ub$. We note that adding $\frac{\lambda_\ell}{n}
\Ib_{d_r}$ to $\Lambdab$ in (\ref{est.cov}) aims to stabilize the
estimator $\widehat\Sigmab_0$, and will not affect its consistency to $\Sigmab_0$.

\begin{rmk}\label{rmk.select_model}
The number $d_r$ in \eqref{sigma} can be used as a guide in
determining how large the model rank is allowed with the given data
size $n$. That is, the value $n-d_r$ should be adequate for error variance
estimation.
\end{rmk}

%Finally, the asymptotic covariance matrix in
%Theorem~\ref{thm.asymp_dist} is estimated by
%\begin{eqnarray}
%\widehat\Sigmab_0=\widehat\sigma^2\widehat\Deltabf_0\left(\widehat\Deltabf_0^T
%\Vb_n\widehat\Deltabf_0+\frac{\lambda_\ell}{n}
%\Ib_{m_p}\right)^{-}\widehat\Deltabf_0^T.\label{est.cov}
%\end{eqnarray}

\section{Multistage Variable Selection for {Genetic Main and G$\bm\times$G Effects}}\label{sec.gg}

By the developed inference procedure of low-rank model, we introduce
in Section~\ref{sec.slr} the SLR-screening. In
Section~\ref{sec.esc}, the SLR-screening is incorporated into the
conventional SC procedure to propose ESC for G$\times$G detection.

\subsection{Sparse and low-rank screening}\label{sec.slr}

Due to the extremely high dimensionality for G$\times$G, a
single-stage Lasso screening is not adequately flexible enough for
variable selection. To improve the performance, it is helpful to
reduce the model size from $m_p$ to a smaller number. The main idea of SLR-screening is
to fit a low-rank model to filter out insignificant variables first,
followed by implementing Lasso screening on the survived variables.
The algorithm is summarized below.\\

%In Stage-1, we utilize the low-rank property of
%$\etab$ and fit a low-rank model; then, based on the asymptotic distribution of
%$\etab(\widehat\phi)$, we use the test statistic
%on each term to screen out significant terms. In Stage-2, we use the survived terms
%to fit a conventional regression model with 1-norm penalty, and
%those
%terms with non-zero estimates are identified. We summarize the procedure below and then explain the details.\\

\hrule

~\

\noindent \textbf{Sparse and Low-Rank Screening (SLR-Screening):}
\begin{itemize}
\item[1.]
\textbf{Low-Rank Screening:} Fit the low-rank model~(\ref{model}).
Based on the test statistics for $\beta_0$, screen out
variables to obtain the index set $\CI_{\rm LR}$.

\item[2.]
\textbf{Sparse (Lasso) Screening:} Fit Lasso on $\CI_{\rm
LR}$. Those variables with non-zero estimates
are identified in $\CI_{\rm SLR}$.
\end{itemize}
\hrule \vspace{0.7cm}

The goal of Stage-1 in SLR-screening is to
screen out important variables by utilizing the low-rank property of $\etab$.
To achieve this task, we propose to fit the low-rank model~(\ref{model}) to obtain $\widehat\beta_{\lambda_\ell}$ and $\widehat\Sigmab_0$.
Based on Theorem~\ref{thm.asymp_dist}, it is then reasonable to screen out variables as
\begin{eqnarray}
\CI_{\rm LR}=\left\{j:\frac{|\widehat\beta_{\lambda_\ell,j}|}
{\sqrt{n^{-1}\,
\widehat\Sigmab_{0,j}}}>\alpha_\ell\right\}\label{I1}
\end{eqnarray}
for some $\alpha_\ell>0$, where $\widehat\beta_{\lambda_\ell,j}$ is
the $j^{\rm th}$ element of $\widehat\beta_{\lambda_\ell}$, and
$\widehat\Sigmab_{0,j}$ is the $j^{\rm th}$ diagonal element of
$\widehat\Sigmab_0$. Here the threshold value $\alpha_\ell$ controls
the power of the low-rank screening.

The goal of Stage-2 in SLR-screening is to enforce sparsity. Based
on the selected index set $\CI_{\rm LR}$, we refit the model with
1-norm penalty through minimizing
\begin{align} \frac{1}{2}\|\Yb-\Xb_{\CI_{\rm LR}}
\beta_{\CI_{\rm LR}}\|^2 +\lambda_{s}\|\beta_{\CI_{\rm
LR}}\|_1,\label{criterion2}
\end{align}
where $\Xb_{\CI_{\rm LR}}$ and $\beta_{\CI_{\rm LR}}$ are,
respectively, the selected variables and parameters in $\CI_{\rm
LR}$, and $\lambda_s$ is a penalty parameter for sparsity
constraint. Let the minimizer of (\ref{criterion2}) be
$\widehat\beta_{\CI_{\rm LR}}$, and define
\begin{eqnarray}
\CI_{\rm SLR}=\left\{j\in \CI_{\rm LR}:\widehat\beta_{\CI_{\rm
LR},j}\neq 0\right\} \label{I2}
\end{eqnarray}
to be the final identified main effects and interactions from the
screening stage, where $\widehat\beta_{\CI_{\rm LR},j}$ is the
$j^{\rm th}$ element of $\widehat\beta_{\CI_{\rm LR}}$. To determine
$\lambda_s$, the $K$-fold cross-validation ($K=10$ in this work) is
applied. Subsequent analysis can then be conducted on those
variables in $\CI_{\rm SLR}$.

%\begin{rmk}\label{why}
%\end{rmk}

%[Why a sparse low-rank model, why not a direct sparse model?]
%Here we provide only some heuristic
%explanations. As for rigorous theoretical justification, it will be
%reported in a separate article.

\subsection{Extended Screen-and-Clean for G$\bm{\times}$G}\label{sec.esc}

\rmSC (SC) of Wasserman and Roeder (2009) is a novel variable
selection procedure. Firstly, the data are split into two parts, one
for screening
%\footnote{The screening part of data can be
%further divided into subparts for cross-validation of the
%screening.}
and the other for cleaning. The main reason of using two independent
data sets is to control the type-I errors while maintaining high
detection power. In the screening stage, Lasso is used to fit all
covariates, of which zero estimates are dropped. The threshold for
passing the screening is determined by cross-validation. In the
cleaning stage, a linear regression model with variables passing the
screening process is fitted, which leads to the LSE to identify
significant covariates via hypothesis testing. A critical assumption
for the validity of SC is the sparsity of effective covariates. As a
consequence, by using Lasso to reduce the model size, the success of
the cleaning stage in identifying relevant covariates is guaranteed.

Recently, SC has been modified by Wu \textit{et al}. (2010) to
detect G$\times$G as described in Section~\ref{sec.1}. This
procedure has been shown to perform well through simulation studies.
However, the procedure can be less efficient when the
number of genes is large.
For instance, there could be many genes
remain after the first screening and, hence, a rather large number
of parameters is required to fit model~\eqref{model_full0} for the
second screening.
As the performance of Lasso depends on the model
size, a further reduction of model size can be helpful to increase
the detection power. To achieve this aim, unlike standard SC that
fits the full model~\eqref{model_full0} with Lasso screening, we
propose to fit the low-rank model~\eqref{model} with SLR-screening
instead. We call this procedure Extended Screen-and-Clean (ECS). Let
$G^*$ be the set of all genes under consideration. Given a random
partition $\mathcal{D}_1$ and $\mathcal{D}_2$ of the original data
$\mathcal{D}$, the ESC procedure for detecting G$\times$G is
summarized below.\\

\hrule

~\

{\noindent\bf Extended Screen-and-Clean (ESC):}
\begin{enumerate}\itemsep=0pt
\item
Based on $\mathcal{D}_1$, fit Lasso on $(Y,G^*)$ to obtain
$\widetilde\xi_{G^*}$ with the 1-norm penalty $\lambda_m$. Let $G$
consist of genes in $\{j:\widetilde\xi_{G^*,j}\neq 0\}$. Obtain
$\E(G)=G\cup\{\mbox{all interactions of $G$}\}$.

\item
Based on $\mathcal{D}_1$, implement SLR-screening on $(Y,\E(G))$ to
obtain $\CI_{\rm SLR}$. Let $\CS$ consist of main and interaction terms in $\CI_{\rm SLR}$.

\item
Based on $\mathcal{D}_2$, fit LSE on $(Y,\CS)$ to obtain estimates
of main effects and interactions $\widehat\xi_\CS$ and
$\widehat\etab_\CS$. The chosen model is
\begin{eqnarray*}
\mathcal{M}=\displaystyle{\left\{g_j,g_kg_l\in
\CS:|T_j|>t_{n-1-|\mathcal{S}|,\frac{\alpha}{2|\mathcal{S}|}},|T_{kl}|>t_{n-1-|\mathcal{S}|,\frac{\alpha}
{2|\mathcal{S}|}}\right\}},
\end{eqnarray*}
where $T_j$ and $T_{kl}$ are the $t$-statistics based on elements of
$\widehat\xi_\CS$ and $\widehat\etab_\CS$, respectively.
\end{enumerate}

\hrule

~\ \\

\begin{figure}[htb]
\includegraphics[width=\textwidth]{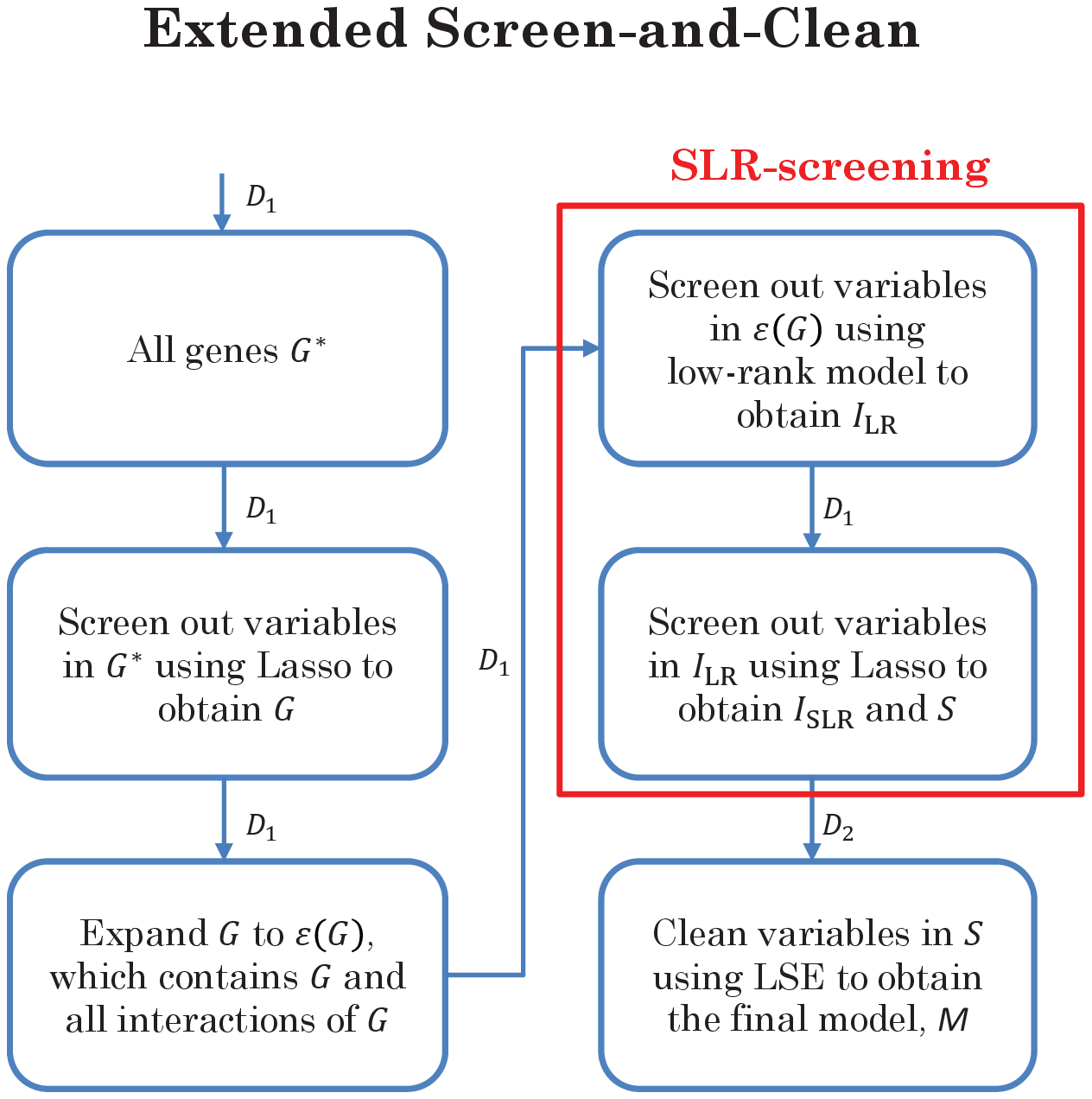}
\caption{Flowchart of ESC for detecting G$\times$G. The arrow
indicates which part of the data is used. The case of SC replaces
SLR-screening by Lasso screening.} \label{Lasso_tensor}
\end{figure}

\noindent For the determination of $\lambda_m$ in Step-1 of ESC, in
Wu \textit{et al}. (2010) they use cross-validation. Later, Liu,
Roeder and Wasserman (2010) introduce StARS (Stability Approach to
Regularization Selection) for $\lambda_m$ selection, and this
selection criterion is adopted in the R code of \textit{Screen} \&
\textit{Clean} (available at {\tt
http:/\!/wpicr.wpic.pitt.edu/WPICCompGen/}). Note that the intercept
will be included in the model all the time.
%In the cleaning stage, the
%intercept will be determined to stay in or to drop out of the final
%model depending on being significant or not.
Note also that the proposed ESC is exactly the same with Wu's SC,
except SLR-screening is implemented in Step-2 instead of Lasso
screening. See Figure~\ref{Lasso_tensor} for the flowchart of ESC.

\section{Simulation Studies}\label{simulation}

Our simulation studies are based on the design considered in Wu et al. (2010) with some extensions.
In each simulated dataset, we generated genotype and trait values of $400$ individuals. For genotypes, we
generated 1000 SNPs, $G=[g_1, \cdots, g_{1000}]^T$ with
$g_j\in\{0,1,2\}$, from a discretization of normal random
variable satisfying $P(g_j= 0) = P(g_j = 2) = 0.25$ and $P(g_j = 1)
= 0.5$. The 1000 SNPs can be grouped into $200$ 5-SNP blocks,
with which SNPs from different blocks are independent and SNPs within the same block are correlated with
$R^2 = 0.3^2$.
Conditional on $G$, we generate $Y$ using the following 4 models, where $\beta$ is the effect size and $\varepsilon \sim N(0,1)$:
\begin{enumerate}\itemsep=0pt
\item[M1:]$Y=\beta(g_5g_6+0.8g_{10}g_{11}+0.6g_{15}g_{16}+0.4g_{20}g_{21}+0.2g_{25}g_{26})+\varepsilon$.% (m12).
\item[M2:]$Y=\beta(g_5g_6+0.8g_{10}g_{11}+0.6g_{15}g_{16}+2g_{20}+2g_{21})+\varepsilon$.% (m6).
\item[M3:]$Y=\beta\vecp(\etab)^T \vecp(\Jb)+\varepsilon$, $\eta_{jk}=0.9^{|j-k|}$ for $1\le j\ne k\le6$ and $\eta_{jk}=0$ for $j,k>6$.% (m10).
\item[M4:]$Y=\beta\vecp(\etab)^T \vecp(\Jb)+\varepsilon$, where we randomly generate $\eta_{jk}={\rm
sign}(u_1)\cdot u_2$ with $u_1\sim U(-0.1,0.9)$ and $u_2\sim
U(0.5,1)$ for $1\le j\ne k\le8$, and $\eta_{jk}=0$ for
$j,k>8$.%(m11).
%\item[M5:]$Y=\beta\vecp(\etab)^T \vecp(\Jb)+\varepsilon$, $\eta_{jk}=1$ for $1\le j\ne k\le10$ and $\eta_{jk}=0$ for $j,k>10$.% (m9).
\end{enumerate}

To compare the performances, let $\mathcal{M}_0$ denote the index
set of nonzero coefficients of the true model, and let $\mathcal{M}$
be the estimated model. Define the \textit{power} to be
$E(|{\mathcal{M}}\cap \mathcal{M}_0|/|\mathcal{M}_0|)$,
% (the fraction of discoveries of
%$\mathcal{M}_0$ over the size of $\mathcal{M}_0$),
the \textit{exact discovery} to be $P({\mathcal{M}}=
\mathcal{M}_0)$,
%(the probability of all recoveries of $\mathcal{M}_0$ without any
%false discovery),
the \textit{false discovery rate} (FDR) to be $E(|{\mathcal{M}}\cap
\mathcal{M}_0^c|/|{\mathcal{M}}|)$,
% (the fraction of false
%discoveries over the size of ${\mathcal{M}}$),
and the \textit{type-I error} to be $P({\mathcal{M}}\cap
\mathcal{M}_0^c\ne\varnothing)$.
%(the probability that at least one
%false discovery).
These quantities are reported with 100 replicates for each model.

\begin{figure}[h]
\centering
\includegraphics[width=3.3in]{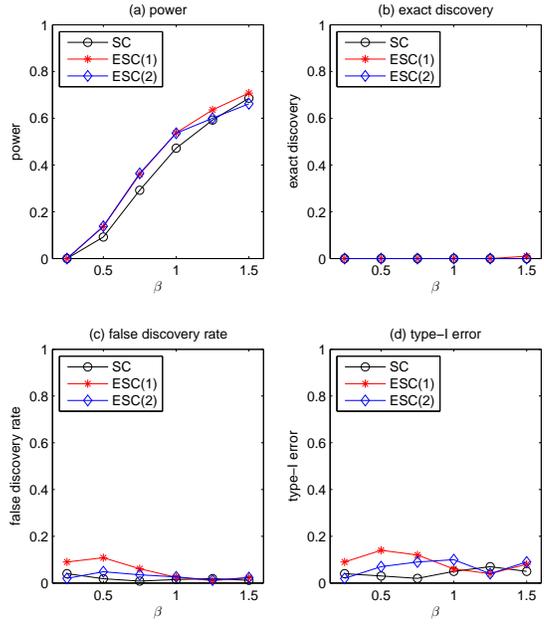}
\caption{Simulation results under M1.} \label{fig.m1}
\end{figure}

\begin{figure}[h]
\centering
\includegraphics[width=3.3in]{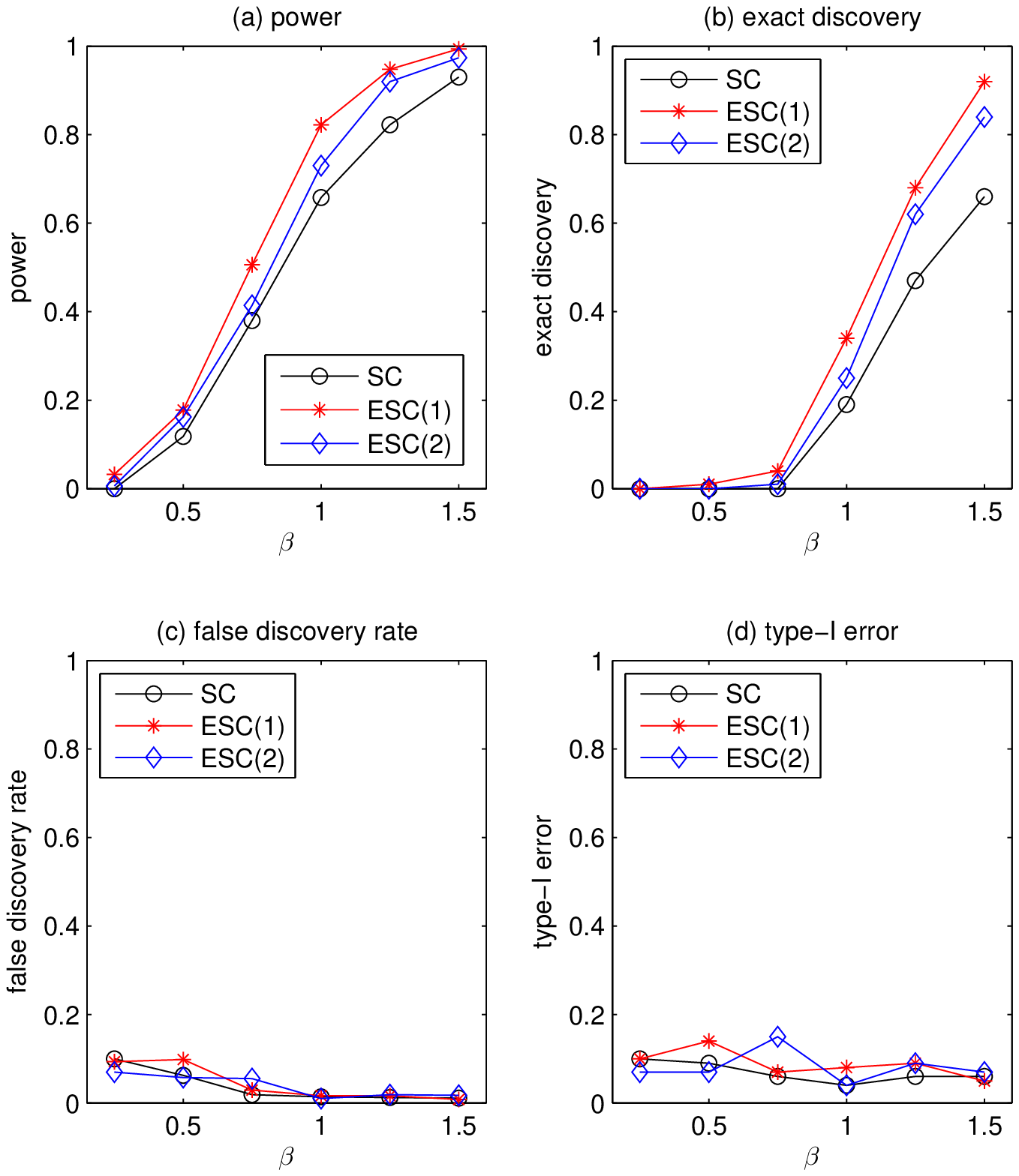}
\caption{Simulation result under M2.} \label{fig.m2}
\end{figure}

\begin{figure}[h]
\centering
\includegraphics[width=3.3in]{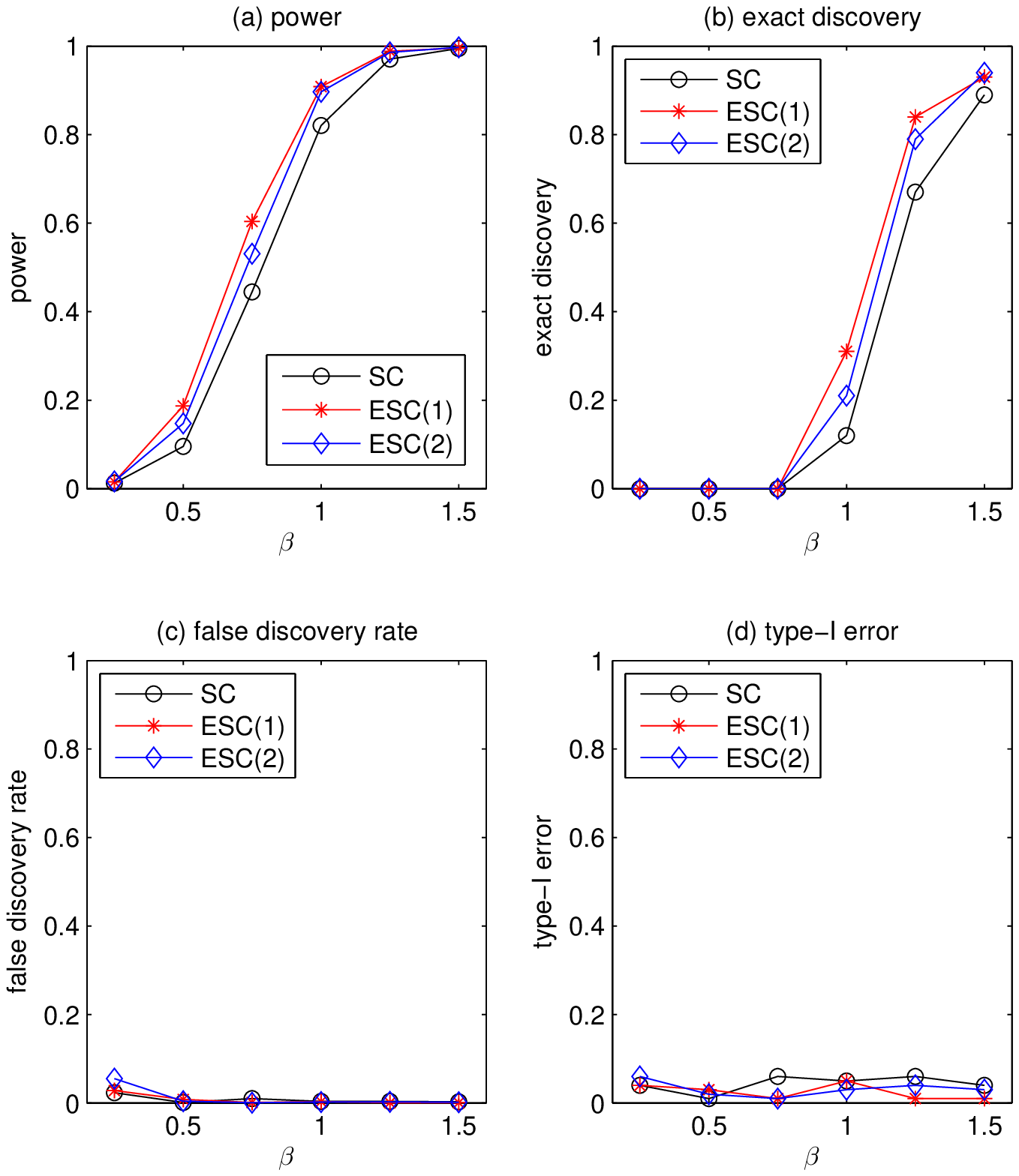}
\caption{Simulation result under M3.} \label{fig.m3}
\end{figure}

\begin{figure}[h]
\centering
\includegraphics[width=3.3in]{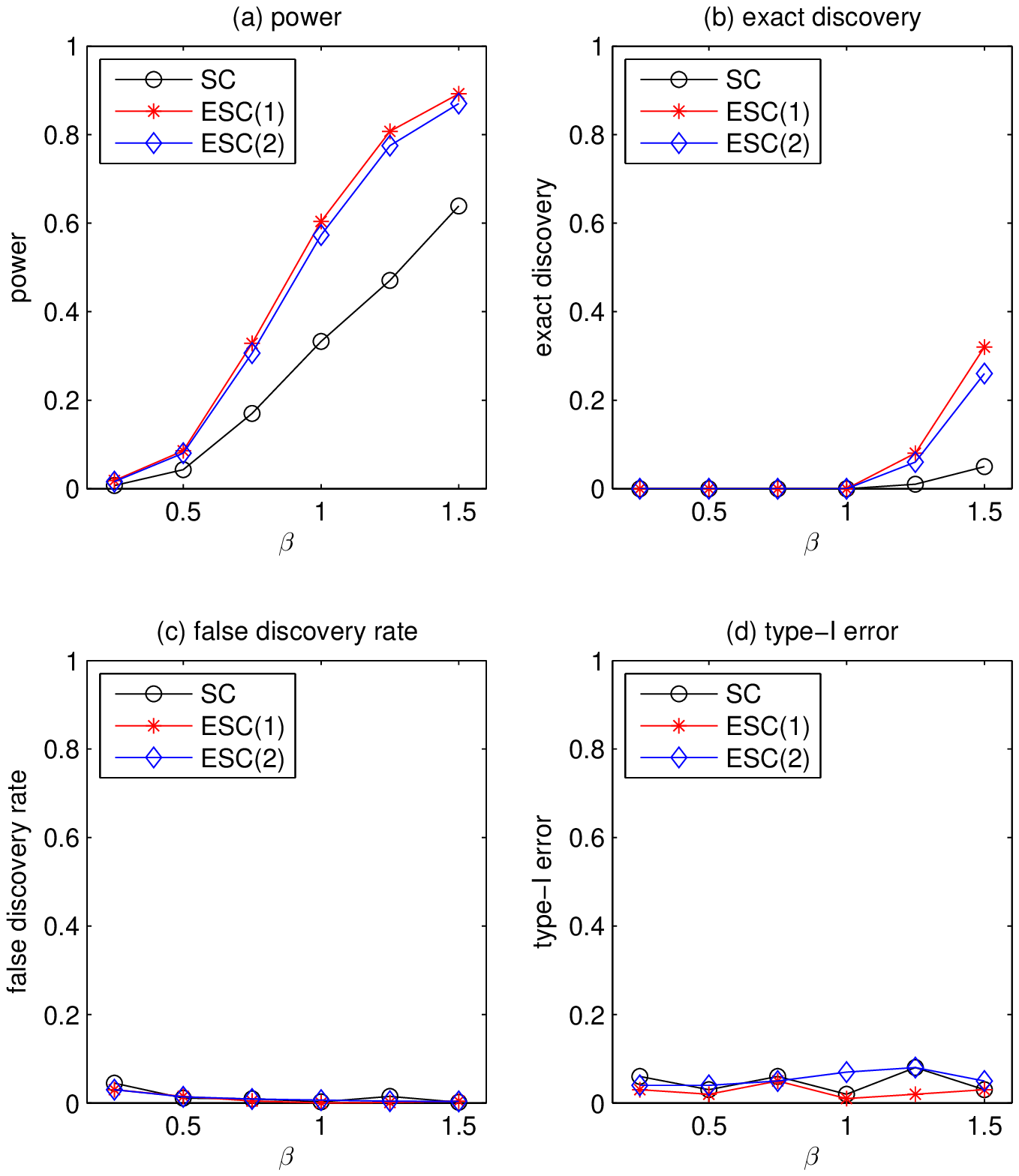}
\caption{Simulation result under M4.} \label{fig.m4}
\end{figure}

Simulation results under different model settings are placed in
Figures~\ref{fig.m1}-\ref{fig.m4}. It can be seen that both ESC(1)
and ESC(2) can control FDR and type-I error adequately in all
settings. In the pure interaction model M1, ESC(1) is
the best performer, while the performances of SC and ESC(2) are
comparable. Interestingly, when the true model contains main effects
(M2, Figure~\ref{fig.m2}), both ESC(1) and ESC(2) do outperform SC
obviously for every effect size $\beta$. It indicates that
conventional SC using model~(\ref{model_full0}) is not able to
identify main effects efficiently. We found SC procedure is more
likely to wrongly filter out the true main effects in the second
Lasso screening stage. However, with the low-rank screening to
reduce the model size, these true main effects have higher
chances to enter the final LSE cleaning and, hence, a higher power
of ESC is reasonably expected.
% Models M3-M4 are designed to mimic
% the situation that a set of genes interact with each other to affect
% the response. From the simulation results in
% Figures~\ref{fig.m3}-\ref{fig.m4},
The superiority of ESC procedure
can be more obviously observed under models M3-M4 (Figures~\ref{fig.m3}-\ref{fig.m4}),
where the powers and exact discovery
rates of ESC(1) and ESC(2) dominate that of SC for every effect size
$\beta$. One reason is that there are many significant interactions
involved in M3-M4, and ESC with a low-rank model is able to correctly
filter out insignificant interactions in $\etab$ to achieve better
performances. In contrast, directly using Lasso screening does not
utilize the matrix structure of $\etab$. On one side, it tends to
wrongly filter out significant interactions. On the other side, it
tends to leave too many insignificant terms in the screening stage.
Consequently, the subsequent LSE does not have enough sample size to
clean the model well, and results in lower detection powers.

We note that although the rank of $\etab$ in models
M1-M4 ranges from 6 to 8, ESC with rank-1 and rank-2 models suffice
to achieve good performances. It indicates the robustness and applicability of the
low-rank model~\eqref{model}, even with an incorrectly specified rank $r$.
Moreover, we observe that ESC(1) outperforms ESC(2) in most of the
settings. Given that the aim of low-rank screening in SLR-screening is
to reduce the model size, a good approximation of $\etab$ is
capable to remove non-important terms.
%% and there is no need to estimate $\etab$ correctly.
In contrast, while the rank-2 model approximates $\etab$ more precisely, it also requires more parameters
in model fitting. With limited sample size, the gain in
approximation accuracy from rank-2 model cannot compensate the loss
in estimation efficiency and, hence, ESC(2) may not have a better
performance than ESC(1) does. See also Remark~\ref{rmk.select_model}
for the discussion of selecting $r$ in ESC procedure.

\clearpage

\end{document}